\documentclass[a4paper]{article}
\usepackage{preamble}

\title{Impact Analysis of the Chesa Boudin Administration}
\author{Jordan G. Taqi-Eddin\\
        \small University of California, Berkeley\\
        \small jgte29@berkeley.edu
}
\date{ }

\begin{document}
\maketitle

\begin{abstract}

Claims of soft-handed prosecutorial policies and increases in crime were precipitating factors in the removal of Chesa Boudin as district attorney of the city and county of San Francisco. However, little research has been conducted to empirically investigate the veracity of these indictments on the former district attorney. Using regression discontinuity design (RDD), I find that the Boudin administration led to a 36\% and 21\% reduction in monthly prosecutions and convictions respectively for all crimes. Moreover, his tenure increased monthly successful case diversions by 58\%. When only looking at violent crimes during this period, the SFDA's office saw a 36\% decrease, 7\% decrease, and 47\% increase in monthly prosecutions, convictions, and successful case diversions respectively. Although, the decrease in monthly convictions was not statistically significant for the violent crimes subset. Additionally, I did identify a potentially causal relationship between lower numbers of prosecutions and higher levels of criminal activity, however, such findings did not meet the standard for statistical significance. Finally, I conclude that using machine learning algorithms, such as neural networks and K-nearest neighbors, in place of ordinary least squares regression for the estimation of the reduced form equation possibly may decrease the size of the standard errors of the parameters in the structural equation. However, future research needs to be conducted in this space to corroborate these initially promising findings. 

\end{abstract}

\setcounter{page}{0}


\begin{multicols}{2}
\section{Introduction}
Chesa Boudin's tenure as district attorney of San Francisco was characterized by turmoil and political strife. This dissension culminated in the summer of 2022 with the recall and removal of Boudin. While initially popular with his constituency when he took office in January 2020, his policies, such as increasing case diversions and sentencing reform, caused him to quickly fall out of favor with the public as levels of crime started to steadily increase. \protect\cite{harvard_law_review} From 2020 to 2021, the San Francisco Police Department saw a 8.7\% increase in incident reports for violent crimes (13,466 to 14,643) and 9.3\% increase for all crimes (117,198 to 128,127). \protect\cite{police_department_2024} Moreover, as a result of the sharp uptick in property crimes and anti-Asian hate crimes, the progressivism that Boudin once was touted for became an indelible mark of his inability to continue as district attorney in the eyes of San Francisco voters. \par

A bipartisan coalition of politicians, community organizations, and investors first set out to oust the district attorney approximately a year into his term, however, to no avail. \protect\cite{harvard_law_review} Following the initial recall campaign, the organizers' second attempt to surpass the requisite amount of roughly fifty-one thousand signatures was successful in October 2021. Securing over eighty-three thousand signatures meant that Boudin and his policies would face trial in the court of public opinion on June 7, 2022. In a resounding defeat, Boudin saw more San Franciscans vote in favor of his removal from office than those who had originally elected him. \par

Voters' vehement rejection of Boudin's time as San Francisco District Attorney exemplifies the fragility of a constituency's trust in elected officials. Moreover, it provides a lens through which crime statistics' complex relationship with governance can be explored. Crime statistics are much more than just data, they additionally uphold an exert agency upon the social domains which they are set to describe. The collection of crime statistics puts a heightened focus on law enforcement, thereby increasing the prominence of police activity which consequently leads to the further creation of data on criminal behavior. \protect\cite{brayne2017big} This negative feedback loop has solidified a strong relationship of co-production between crime and crime statistics. \par

Often, the consequence and power that comes with gathering and disseminating crime statistics is overlooked due to the perceived impartiality of such data. \protect\cite{muhammad2019condemnation} It is regarded as a tool through which individuals can observe some objective truth. However, within this notion arises a glaring paradox. Data is used as a tool due to being devoid of bias, but those who use it are not of the same disposition. Individuals possess convictions, motives, and virtues which dictate how they comport themselves. Thus, data, and crime statistics by that effect, are subject to these very forces as well. \par


In the sphere of American political discourse, there has been a long history of data being weaponized to preserve systems that perpetuate inequality. \protect\cite{muhammad2019condemnation} From upholding the discriminatory practices of Jim Crow to pushing back against reforms to increase access to education for African Americans, crime statistics are an extremely effective way for those in power to justify unjust social structures and hinder progress through fear. \par

Given the fact that the increase in crime incidents was a catalyst for the recall campaign of Chesa Boudin, it poses the question of whether the early end to his administration was a consequence of the aforementioned phenomena. There undoubtedly was an increase in SFPD incident reports from 2020 to 2021, however, to properly identify the underlying factors for this rise in crimes a more nuanced analysis is required than solely looking at who was the district attorney during this period and attributing culpability accordingly. In addition to ushering of the Boudin administration in 2020, another event had a monumental impact on nearly every aspect of life in San Francisco, the start of the COVID-19 pandemic. In the year prior to the pandemic, San Francisco had an approximately 4.5\% decrease in violent crime incident reports (16,821 to 16,066) and a 3.2\% decrease when factoring in all crimes (151,672 to 146,847). \protect\cite{police_department_2024} When looking at incident reports for all crimes, the year-to-year decrease in reports from 2018 to 2019 is very similar to that from 2022 to 2023, which was roughly 3\% (135,363 to 131,260). The number of violent incident reports from 2022 to 2023 still increased by around 0.1\% (15,522 to 15,542), but the magnitude of these year-to-year increases has gradually been falling from the 8.7\% spike seen between 2020 and 2021. The fact that these rates resemble pre-COVID levels indicates that the pandemic may have contributed to the irregular trends in crime levels in San Francisco. In this paper, I set out to answer whether the drastic increases in criminal activity in 2020 were solely a consequence of Boudin's policies, or additionally a byproduct of crime rates returning to homeostasis.

\section{Literature Review}

Through the following sections of this paper, I set out to identify any causal relationship between Boudin and the increases in crime during his tenure. Consistent with prior work completed to analyze the impact of federal regulation in other domains, policy is commonly used as an instrumental variable when trying to establish causality. \protect\cite{deschenes2017defensive} In their work investigating the causal impact of the U.S. $NO_{x}$ budget program on health, Dr. Deschenes and his team use the implementation of the policy in a region of the country as an instrument for the area's ozone concentration. While their paper was interested in federal regulation within the environmental conservation space, it is still relevant to my work as it demonstrates the power of using policy as an instrumental variable when attempting to identify causality. Furthermore, previous literature demonstrates that research discontinuity design in concert with instrumental variables can be an effective strategy for investigating causal relationships. \protect\cite{robinson2011evaluating} \protect\cite{angrist1999using} This "fuzzy regression-discontinuity" design is especially helpful in instances where there is not data for both the treatment and control groups both before and after the intervention, which is requisite to use a difference in difference model. \protect\cite{card1993minimum} \par

In addition to conventional quasi-experimental methods, there are numerous additional strategies commonly incorporated into the research design of impact analyses on policing and legal reform. Trends in crime rates can be estimated using incident reports from police departments. \protect\cite{kahmann2022impact} Furthermore, by analyzing the subset of crime data associated with violent crimes against all crimes, potential disparate impacts of policies for violent crimes can be pinpointed. In some literature, lags are used to capture the notion that policies may not have immediate impacts, and also have the potential to have residual effects years after they are enacted or terminated. \protect\cite{hernandez2019criminal}

The rise in the prevalence and influence of machine learning has re-imagined the landscape of academic research. However, progress on this front in the social sciences has been quite slow due to one major issue - interpretability. \protect\cite{athey2019machine} As a result of the nature of machine learning algorithms, like neural networks, there is a tradeoff between computational power and inference, a cornerstone of literature in economics and statistics. \protect\cite{efron2022computer} Recently, groundbreaking work has been conducted to improve the interpretability of random forests \protect\cite{aria2021comparison} and neural networks all while preserving the predictive prowess of these models. \protect\cite{wang2020deep} Furthermore, machine learning algorithms, such as maximum likelihood estimators, being combined with instrumental variables, has provided a new way to model causal effects. \protect\cite{maydeu2019instrumental} Such work demonstrates the promise of machine learning's emergence as an insightful complement to conventional empirical analysis techniques.

\section{Data}
For my analysis, I rely on three different data sources --- SFDA case actions data \protect\cite{district_attorney_cases}, SFPD incident reports data \protect\cite{police_department_2024}, and San Francisco neighborhood geospatial data \protect\cite{analysis_neighborhoods}. Below I discuss data usage, feature engineering, and data granularity. \par

3.1.1 \textit{SFDA Case Actions - Data Generation}. The San Francisco case actions dataset was created and is managed by the city's open data portal, DataSF. When the SFDA files charges against a defendant, pertinent data is manually inputted into the District Attorney Office's case management system. On an approximately weekly basis data reports are extracted from this system, subjected to cleaning and anonymization processes, and subsequently incorporated into the portal. \par

The dataset contains information on over 100 thousand cases presented to the District Attorney’s Office from January 2014 to the present day. The dataset only includes cases where the office initiated prosecution via filing new criminal charges or submitting a motion to revoke probation or parole (MTR). Cases handled by the San Francisco Adult Probation Department or the state Division of Adult Parole Operations initiating a motion to revoke are excluded. \par

When the SFDA initiates a new criminal charge, the most serious offense type is classified as the primary offense for which the SFDA prosecutes the defendant. In MTR cases, the most serious offense is categorized based on the initial prosecution, as filing an MTR indicates a pursuit of a new sanction within the sentence of a prior criminal conviction rather than filing new charges for the latest offense. The filing date for MTRs is determined by the date when the SFDA filed the MTR for the new arrest. \par

3.1.2 \textit{SFDA Case Actions - Data Granularity \& Temporal Aggregation}. Each observation in the San Francisco case actions dataset represents a specific action taken by the district attorney's office on a specific case for a defendant. I limit the temporal scope of my analysis to crimes with arrest dates from 1 January 2014 to 31 December 2023. Given the fine granularity of the data, I aggregate the data monthly. The aggregations are performed on the arrest dates, not on the dates of SFDA actions, as such information is not available in the dataset. \par

Before temporal aggregation, I filter the prosecution dataset to independently analyze crime and SFDA action subsets (e.g. Convictions on Violent Crimes). Using the definition put forth by the state of California, crimes in the case actions dataset categorized as violent crimes include --- Assault and Battery, Assault, Robbery, Other Sex Law Violations, Weapons Charges, Hit-and-Run, Homicide, Attempted Homicide, Kidnapping, Arson, Rape, Lewd or Lascivious Behavior, Manslaughter, Vehicular Manslaughter, Child Molestation, Obscenity, Other Enhancements. \protect\cite{california-penal-code-667.5} Another subset I investigate is all crimes except quality-of-life crimes. Offenses that fall into this category are --- Marijuana Possession, Liquor Laws, Disturbing Peace, Disorderly Conduct, Malicious Mischief, Trespassing, Prostitution, Petty Theft, Burglary, and Vandalism. \protect\cite{harvard_law_review} \protect\cite{nordberg2002jails} \protect\cite{california-penal-code-594} \par

3.1.3 \textit{SFDA Case Actions - Feature Engineering}. I log-normalize all of my endogenous variables within the Case Actions dataset --- Prosecutions, Successful Case Diversions, \& Convictions --- to smooth out the trends and improve the linearity of the data. Moreover, I obtain my analysis's synthetic treatment and time-fixed effects via various methods. The simplest feature to develop was the monthly time-fixed effect because it was a byproduct of the temporal aggregation process. However, obtaining the time-fixed effect of the COVID-19 pandemic and the indicator of my treatment, Chesa Boudin being in office, required a more nuanced approach. To account for the COVID-19 pandemic, I use two different strategies. The first one is a simple cutoff at March 2020 for when Shelter-In-Place Orders first went into effect in California. \protect\cite{friedson2021shelter} The second approach uses a range from March 2020 to the end of 2021  to define the time-effect of the pandemic. This range follows literature that concludes that COVID entered its endemic phase by the end of 2021. \protect\cite{ioannidis2022end} \par

As previously mentioned, the dates of SFDA actions are not available via the case actions dataset. Therefore, to approximate the date during which such actions took place, I used a lagging strategy. The average Median Number of Days from Arrest to Conviction for convictions with arrest dates between 2020 to 2022 was approximately 307 days (~10 months) for all crimes and 426 days (~14 months) for the violent crimes subset (See \textit{Appendix A}). This means that for cases with arrest dates towards the end of his tenure, it is highly unlikely that any conviction or successful diversion would have occurred during his administration. Thus, the cutoffs for case actions under the Boudin administration are lagged by 10 and 14 months in the all crimes and violent crimes datasets respectively (i.e. The range for a Boudin conviction for all crimes will be crimes with an arrest date between March 2019 \& September 2021). The same lagging strategy is applied to the violent crimes subset and when analyzing successful case diversions. However, when analyzing new charges filed by the SFDA's office, I do not employ lagging to reflect the fact prosecutors normally file charges not long after the date of arrest, often within 48 hours. \protect\cite{cacourts} So for arrests that took place during a month when Boudin was in office, I classified those as Boudin prosecutions. \par

\textit{Figures 1 \& 2} demonstrate the significant impact of lagging on the overall data trends. By lagging, the mean monthly log total SFDA conviction and successful diversion trends during the Boudin administration --- 2020 to 2022 without lags and 2019 to 2021 with lags --- completely change. Furthermore, lagging causes a massive shift in the pre-Boudin era trend for successful diversions. In addition to epitomizing the power of lagging, the aforementioned figures show that the exclusion of cases with an arrest date in 2023 changes the trends as well. Examining the \textit{Without Lags} plots in \textit{Figures 1 \& 2}, there appears to be a massive drop in log convictions and successful diversions starting from August 2023 onwards. Such declines can be explained by the fact that convictions and case diversions do not occur immediately after a defendant is arrested meaning that many of these cases were still open at the end of my analysis period, 31 December 2023. Therefore, to prevent the temporal scope from introducing bias, I conduct my analysis once using cases from 2014 to 2023 and then again excluding 2023 arrests. For consistency, I additionally employ this control during my analysis of SFDA prosecutions. However, as \textit{Figure 3} shows, the exclusion of 2023 does not cause any major change in the post-Boudin trend. Additionally, the figure shows why using lagging for the prosecutions analysis would not be a good fit. While the pre- and post-Boudin trends are virtually unaffected by the lags, the trend for Boudin's time in office does see a noticeable increase which could have biased the analysis results. \par

3.2.1 \textit{SFPD Incidents - Data Generation}. The dataset comprises incident reports submitted since January 1, 2018. These reports are logged either by law enforcement officers or through self-reporting by the public, utilizing the online reporting system of the San Francisco Police Department. It should be noted that with self-reporting non-response bias has the potential to impact the credibility of the data. For example, recent academic literature indicates that undocumented immigrants under-report crimes due to the fear of potentially facing deportation. \protect\cite{comino2020silence} Therefore, due to the dissimilar demographics of different areas and varying reporting behaviors of social groups, failure to account for this heterogeneity could bias the results of my findings. However, I chose not to employ any control for non-response bias because I am comparing the same communities, those in San Francisco, across different periods in time, not two different regions. So I believe that my spatial aggregation is sufficient in eliminating the aforementioned heterogeneity and avoiding this form of bias. The reports are classified into distinct categories based on the method of report submission and the nature of the incident. These categories include initial reports, which represent the first filing for a given incident; coplogic reports, referring to reports submitted online by the public; and vehicle reports, encompassing incident reports associated with stolen and/or recovered vehicles. \par

Updated daily, data is incorporated into the open dataset only after undergoing a review and approval process by a supervising Sergeant or Lieutenant. Removal of incident reports from the dataset may occur due to compliance with court orders to seal records or for administrative reasons, such as ongoing internal affairs investigations or criminal inquiries. \par

3.2.2 \textit{SFPD Incidents - Feature Engineering}. Much like the SFDA case actions, the SFPD incident reports dataset has fine granularity that requires the usage of temporal aggregation monthly. Additionally, to conduct a chi-squared test of independence, I aggregate yearly across different San Francisco neighborhoods so that I can see if there was any difference in the neighborhood concentrations of crime reports in San Francisco during Boudin's time as District Attorney. To add the month and COVID time-fixed effects, I employ the same strategy I use when creating these features for the SFDA case actions dataset. However, I opt not to use lagging for the synthetic treatment and create the feature in the same manner as when analyzing prosecutions in the case actions dataset. \par

3.3 \textit{Analysis Neighborhoods}. This dataset was created by assigning 2010 Census tracts to neighborhoods based on existing definitions by the Planning Department and the Mayor’s Office of Housing and Community Development. Since 2010, these tracts have not changed. A qualitative assessment determines each tract's appropriate neighborhood based on population distribution and landmarks. The dataset is static. Changes to analysis neighborhood boundaries are evaluated by the Analysis Neighborhood working group, led by DataSF and the Planning Department, with input from various other city departments, and updates are made as needed. I only use the geospatial data in this dataset when creating my neighborhood year-to-year crime prevalence plots.

\section{Methodology}
I use a regression discontinuity design to identify a causal relationship between Chesa Boudin's tenure as SFDA and prosecutions, convictions, and successful diversions. I then build upon the quasi-experimental design and use Boudin's administration as an instrumental variable when investigating the impact of prosecutions on crime levels. Consistent with literature such as \textit{Robinson, 2011} and  \textit{Angrist \& Lavy, 1999}, I use the following regression discontinuity design model: \par

\noindent
(1) $Y_{m,t} = \beta_{0} + \beta_{1} \times \mathds{1}(Boudin)_{m,t} + \mu_{m} + \pi_{m,t} + \epsilon_{m,t}$

\noindent
Here, \textit{m} represents the month and \textit{t} year. The endogenous variables for this model, $Y_{m,t}$, are prosecutions, convictions, and successful diversions, and they are all log-normalized. The synthetic treatment for the model is $Boudin_{m,t}$; it is a binary indicator for whether or not, Chesa Boudin was district attorney during month \textit{m} and year \textit{t}. I lag this variable when analyzing convictions and successful diversions, but not when looking at prosecutions. \par

Examining \textit{Figures 1 --- 3}, it is evident that the outcome variables of my model covary with the month, being most pronounced for prosecutions. To control for this covariate, I include a time-fixed effect, $\mu_{m}$. Moreover, I add a two-way month-by-year fixed effect for the COVID-19 pandemic. As previously noted, I define the COVID-19 pandemic in two different manners --- a simple cutoff at the start of the California Shelter-In-Place-Orders in March 2020 $(Covid_{1})$ \& a range from March 2020 to the end of 2021, when the virus became endemic $(Covid_{2})$. \par

In my instrumental variable approach, I adopt a model similar to that in Hernandez, 2019. \protect\cite{hernandez2019criminal} A prevalent criticism of Boudin during his term was that his policies caused the rise in crime that was seen in San Francisco. To test the merit of this claim, I examine how crime levels in a given year were impacted by the number of prosecutions in the prior year under the two-stage-least squares model below: \par

\noindent
(2) $\hat{\rho}_{m,t-1} = \beta_{0} + \beta_{1} \times \mathds{1}(Boudin)_{m,t-1} + \mu_{m} + \pi_{m,t-1} + \epsilon_{m,t-1}$ \par

\noindent
(3) $Y_{m,t} = \alpha_{0} + \alpha_{1} \times \hat{\rho}_{m,t-1} + \mu_{m} + \pi_{m,t} + \eta_{m,t}$ \par

\noindent
For the reduced form equation, $\rho$ is the log prosecution totals in month \textit{m} during year \textit{t}. I use the same fixed effects and exogenous variables as in the research discontinuity design. However, given the fact that the target variable in equation (2) is log prosecutions, I do not lag the Boudin binary indicator. \par

To investigate whether or not there was any difference in the distribution of crime reports in San Francisco during Boudin's time as District Attorney (i.e. Creation of crime "hotspots") I perform a chi-squared test of independence using contingency tables. For the sake of simplicity in the analysis, I compare the counts during 2019 to the counts during 2021. The reason for this is that the counts are aggregated yearly, so I want to compare the most recent pre-Boudin year to the only complete year during which Boudin was district attorney. The distribution of incident counts amongst neighborhoods is not normal. However, as noted in \textit{McHugh, 2013}, normality is not required for the test and can be performed on distributions where the data is "seriously skewed or kurtotic." \protect\cite{mchugh2013chi} Furthermore, I choose to conduct the tests on the distribution of the concentration of crimes in the neighborhoods during the years in question. I define concentration in the following manner:

\noindent
(4) $Concentration_{n,t} = \frac{\textit{Incident Count}_{n,t}}{\textit{Total Incidents}_{t}};$

$\forall \hspace{0.1 cm} \text{Neighborhoods } n, \text{ during time } t$

As previously noted, there was a stark difference in the number of crimes during 2019 and 2021, as seen in \textit{Figure 4}. Thus, failing to address this, would have drastically biased my results. So to control for this potential source of bias, I look at the proportions of yearly incident reports that come from each neighborhood in place of the raw counts.

Advances in machine learning have caused a paradigm shift in academia. Over the past few years, massive efforts have been made to integrate such algorithms into economic research. \protect\cite{athey2019machine} As a recognition of this substantial progress, I explore new ways to intertwine the spaces of econometrics and machine learning. Using $\textbf{Z}_{i}$ to reference the matrix with the instrumental variables, the two-stage-least-squares estimator is \par

\noindent
(5) $\hat{\textbf{X}}_{i} = \textbf{Z}_{i}^{T}[\textbf{Z}_{i}^{T}\textbf{Z}_{i}]^{-1}\textbf{Z}_{i}^{T}\textbf{X}_{i}$ \par

\noindent
(6) $\hat{\textbf{B}}_{i} = [\hat{\textbf{X}}_{i}^{T}\hat{\textbf{X}}_{i}]^{-1}\hat{\textbf{X}}_{i}^{T}\textbf{Y}_{i}$ \par

\noindent
where $\hat{\textbf{X}}_{i}$ is the fitted values of the data $\textbf{X}_{i}$, $\textbf{Z}_{i}^{T}\textbf{X}_{i} \neq 0$, \& $\textbf{Z}_{i}^{T}\epsilon_{i} = 0$. \protect\cite{madansky1964instrumental} In addition to the conventional usage of generalized least squares to compute the 2SLS estimator, I use neural networks and k-nearest-neighbor algorithms to calculate $\hat{\textbf{X}}_{i}$. Prior research has explored the usage of ML models such as maximum likelihood estimators \protect\cite{maydeu2019instrumental}. However, this work has focused on the structural equation, and does not address a major current issue with using machine learning in economic research, the limited inference from the algorithms. These frequentist approaches already provide coefficients and standard errors that can be easily interpreted, something that is not currently as simple with models such as neural networks. \protect\cite{wang2020deep} \par 
In my research, I attempt to realize the advantages of both deep learning algorithms and generalized least squares. By strategically using the ML model in the reduced form equation but continuing to use generalized least squares for the structural form, I aim to exploit the computational sophistication of these algorithms to obtain reduced standard errors all while keeping the interpretability that is possible with traditional 2SLS (Refer to \textit{Appendix B} to see how the performance of the machine learning algorithms fares against the GLS estimations of the reduced form equation).

\section{Results}
In the following subsections, I report the estimated impacts of the Chesa Boudin administration on SFDA prosecutions, convictions, and case resolutions. Moreover, I detail the effect of the former district attorney on crime levels in San Francisco during his time in office. Lastly, I compare the performances of the two-stage-least squares models that used neural networks and k-nearest-neighbors algorithms to compute the first-stage equation against that of conventional generalized least squares regression.

\noindent
4.1 \textit{Case Actions} \textit{Panel A} of \textit{Tables 1 --- 3} are demonstrative of the significant effect that Chesa Boudin had on total monthly prosecutions. The Boudin administration caused approximately a 31\% drop in monthly prosecutions for all crimes, and a 29\% and 25\% reduction for the violent crimes and no quality-of-life crimes subsets respectively. Furthermore, all these decreases are statistically significant at a $p = 0.01$ level using either $Covid_{1}$ or $Covid_2$, and both inclusive and exclusive of crimes with an arrest year of 2023. However, as previously noted, the exclusion of 2023 for the analysis of protection is unnecessary because I do not lag the $Boudin$ variable, and is included solely for the sake of consistency with the analyses of convictions and case diversions in which the lagging is employed. \par

Due to the fact that the enactment of California's Shelter-in-Place orders in March 2020 compelled San Franciscans to stay indoors, it is not a shock that there was such a considerable dip in the number of prosecutions. This notion is corroborated by \textit{Figure 4} which shows a tremendous drop in the total SFPD incident reports for 2020, a 20\% decline (146,847 to 117,198) for all crimes and a 16\% reduction (16,066 to 13,466) for the violent crimes subset. \par

The conclusions of the prosecutions analysis are clear and consistent across all the different combinations of controls and data subsets. However, when looking at SFDA convictions, the results are not as definitive. For the all crimes and no quality-of-life crimes subsets, the Boudin administration led to a roughly 36\% and 21\% drop in convictions with statistical significance at a $p = 0.01$ level. A different story is seen for the violent crimes subset in \textit{Table 2} as the effect of Boudin is negative, but not statistically significant. The reduction in convictions during the Boudin administration is matched by a significant increase in successful case diversions, as shown in \textit{Panel C} of \textit{Tables 1 --- 3}. Boudin caused an approximately 58\% increase in case diversions for all crimes, 47\% for violent crimes, and 69\% for quality-of-life exclusive crimes. Additionally, these increases were all statistically significant at a $p = 0.01$ level. These upticks in successful diversions fall in line with the promises of criminal justice reform that Boudin avowed during his initial bid to become district attorney. \par

\noindent
4.2 \textit{Incident Reports} After identifying the strong causal relationship between Boudin and the decrease in prosecutions, rise in successful diversions, and to a somewhat lesser extent the drop in convictions, I set out to determine whether these changes in SFDA action trends led to a rise in crime levels throughout the city. In my two-stage least-squares setup, I use Boudin's presence in office as an instrumental variable for the number of monthly prosecutions. Since I establish that Boudin caused a significant drop in prosecutions through my research discontinuity design, I determine whether his tenure caused an increase in crime by seeing if prosecutions are a deterrent for crime (i.e. If an increase in prosecutions during month \textit{m} in year \textit{t-1} causes a decrease in incidents in month \textit{m} during year \textit{t}). \par 

Under all three of the two-stage least-squares models I use, the conclusions of Boudin's impact on crime levels appear to be completely dependent on how I define the COVID-19 pandemic. Regardless of the set of assumptions for COVID that are used for a given regression, an increase in monthly prosecutions in year \textit{t-1} overall seems to lead to a drop of SFPD monthly incidents in year \textit{t}, indicative of prosecutions potentially being a deterrent of crime. In columns (1), (3), \& (5) of \textit{Tables 5 --- 7}, the decreases in monthly incidents associated with a percent rise in prosecutions ranges from about 0.33 to 0.34\% for all crimes, 0.30 to 0.31\% for violent crimes, and 0.18 to 0.20\% for quality-of-life exclusive crimes. However, as columns (2), (4), \& (6) of these tables show, the decreases are only statistically significant when using $Covid_{1}$, as none of the values for the effect of $\widehat{\text{Log Prosecutions}}_{t-1}$ as statistically significant in the regressions that have the $Covid_{2}$ control. \par

In addition to the inconclusive evidence over whether Boudin's administration caused the increase in crime observed during his term, there did not seem to be a distributional change in crime levels throughout San Francisco. \textit{Figures 5 \& 6} indicate that certain neighborhoods, such as the Mission District, in San Francisco have immensely higher prevalences of crimes than others (Refer to the map of the neighborhoods of San Francisco in \textit{Appendix C}). However, as the results from the chi-squared test of independence in \textit{Table 4} show, there was no "pooling" of crime in these neighborhoods during the Boudin administration. Hence, the general neighborhood crime concentrations before and after Boudin's time as district attorney are relatively similar, albeit at much lower levels now. 

\noindent
4.3 \textit{Machine Learning 2SLS} The final part of my assessment pertains to the performance of my ML-GLS 2SLS hybrid models against traditional 2SLS. With structural equation estimations comparable to those using traditional 2SLS, the results of the 2SLS hybrid models are promising. Although there is not a noticeable improvement in the standard errors of the neural network-GLS and knn-GLS 2SLS models over solely using GLS to compute the first- and second-stage equations. Furthermore, for a few of the regressions, not only are the standard errors larger for the ML hybrids, but the machine learning algorithms also had higher first-stage MSEs than GLS (See \textit{Appendix B}). The mixed results indicate that the improvements or worsening in MSEs and standard errors that I obtained in my analysis potentially are the byproduct of random noise. Therefore, while the results from my proposed ML-GLS hybrid 2SLS model are encouraging, much more research in this space will need to be conducted to reach a definitive conclusion on the merits of such an approach.

\section{Discussion}
In the expansive realm of San Francisco's criminal justice landscape, my comprehensive analysis navigates through intricate webs of prosecutorial decisions, crime dynamics, and innovative methodological approaches. The multifaceted nature of my findings prompts a nuanced and extensive exploration of their implications and contributions to the broader academic discourse. The results of my study indicate that Chesa Boudin did have a significant impact on levels of case actions during his time as district attorney of San Francisco. Moreover, it appears that there is a potential link between lower convictions and higher crime levels. Although my research does not enable me to come to a definitive conclusion. \par

The results from my research discontinuity design analysis signal a paradigm shift in the city's prosecutorial strategies during his time in office. Monthly prosecutions and convictions respectively dropped by 31\% and 21\% for all crimes, 29\% and 7\% for violent crimes, and 25\% and 21\% for quality-of-life exclusive crimes. Conversely, successful case diversions rose by  58\%, 47\%, and 69\% respectively. The decreasing prosecutions and convictions in conjunction with increased case diversions underscores Boudin's effort to enact criminal justice reforms in San Francisco.
The overall statistical robustness of these results, even when accounting for the intricate temporal overlaps with the COVID-19 pandemic, solidifies their credibility. However, the challenge of disentangling the effects of Boudin's administration from the broader impacts of the pandemic should be noted. \par

My instrumental variable approach, utilizing Boudin's presence as an instrument for monthly prosecutions, delves into the relationship between law enforcement actions and subsequent crime levels. From the results of my analysis, I cautiously report that the drop in prosecutions caused by Boudin did not have a causal effect on the rising levels of crime that occurred during his term. There are a few reasons why I am unable to present my conclusions with more conviction. Firstly, the statistical significance being contingent on the way in which I define the pandemic is potentially problematic. Moreover, the Boudin administration did not only have an impact on prosecutions. In my research, I demonstrate that Boudin had a causal effect on convictions and case diversions. Thus, Boudin's impact on crime levels may not solely be through prosecution levels, which would be a violation of the exclusion restriction for a good instrument.
The lack of a clear causal relationship between Boudin and rising crime rates despite such a connection being a driving factor behind his removal from office is demonstrative of criminal statistics' power to engender public fear and outcry for political expediency. \protect\cite{muhammad2019condemnation} \par

Integrating machine learning into my two-stage least-squares models introduces a methodological frontier with promising potential. The ML-GLS hybrid 2SLS models exhibit comparable results to traditional methods, opening a pathway for enhanced predictive accuracy. The mixed findings in standard errors and mean squared errors (MSEs) prompt further inquiry, urging researchers to delve deeper into the intricacies of machine learning's role in advancing econometric methodologies. While the results showcase promise, the nascent nature of this approach necessitates ongoing exploration and refinement. \par

My study grapples with inherent limitations, foremost among them the challenge of isolating the unique impact of Boudin's tenure from the concurrent effects of the COVID-19 pandemic. The inconclusive evidence regarding the potential increase in crime during Boudin's term underscores the methodological complexities inherent in attributing causality within the fluid dynamics of urban crime. Future research endeavors should prioritize the refinement of control strategies, explore alternative methodologies, and incorporate neighborhood-specific analyses to unravel the intricate threads of influence. \par

One potential avenue of future exploration would be the implementation of a difference and difference model. As previously noted, I could not perform a difference-in-differences model as I did not possess data for both the treatment and control groups both before and after treatment. Therefore, additional research, gathering data on similar counties in the Bay Area, such as Alameda, or more distant metropolitan regions of comparable size, like Los Angeles, would enable a DID model to be used. Furthermore, collecting data at the neighborhood model would open up the door to implement Heckman similarity matching as well. \protect\cite{heckman1998matching} \par

Another space for additional inquiry would be to examine how additional criminal justice policies other than just prosecutions serve as a deterrent or creator of crime. There is various literature that has focused on the impact of bail \protect\cite{zhou2021empirical} and sentencing \protect\cite{lofstrom2020effect} reform on crime. Research dedicated to individually investigating the impact of similar policies in San Francisco could help better understand the dynamic and complex nature of justice reform in the city. Additionally, it would add another layer of depth to the analysis of Boudin's effect on crime and better differentiate his impact from other outside forces.

\section{Conclusion}
In the tapestry of San Francisco's criminal justice narrative, our study contributes not just as an empirical inquiry but as a catalyst for deeper reflections and ongoing investigations. The intricate dance between prosecutorial strategies, crime dynamics, and innovative methodologies invites scholars to journey further into the heart of urban justice, exploring avenues that transcend the confines of conventional analyses. As we conclude this exploration, the threads of inquiry continue, weaving a narrative that extends far beyond the scope of our study, fostering a deeper understanding of the intricate relation between law, order, and societal dynamics. \par

The study's contributions extend beyond empirical observations, providing a nuanced lens through which to examine the interplay of prosecutorial decisions and crime dynamics in an urban context. The shifts in convictions and successful diversions align with the reform-oriented agenda espoused by Boudin, emphasizing the need for a holistic understanding of criminal justice outcomes. My work invites researchers to delve deeper into the evolving landscape of criminal justice policies, exploring alternative causal pathways and refining methodological approaches. \par

In the synthesis of these diverse facets, my study becomes a testament to the complexity inherent in urban criminal justice systems. The interwoven threads of prosecutorial decisions, law enforcement dynamics, and methodological innovations form a rich tapestry that beckons further exploration. Beyond the immediate implications for San Francisco, our work prompts broader questions about the adaptability of traditional econometric methods and the evolving nature of justice systems in contemporary urban environments. \par

\end{multicols}

\newpage
\bibliographystyle{acm}
\bibliography{sample-base}

\begin{thebibliography}{10}

\bibitem{analysis_neighborhoods}
Analysis neighborhoods.
\newblock [Dataset], 2023.

\bibitem{harvard_law_review}
San francisco district attorney chesa boudin recalled.
\newblock {\em Harvard Law Review Vol. 136}, No. 6 (2023), 1740--1747.

\bibitem{cacourts}
How criminal cases work, 2024.

\bibitem{california-penal-code-594}
California penal code § 594, year.
\newblock \S 594.

\bibitem{california-penal-code-667.5}
California penal code § 667.5, year.
\newblock \S 667.5.

\bibitem{angrist1999using}
{\sc Angrist, J.~D., and Lavy, V.}
\newblock Using maimonides' rule to estimate the effect of class size on scholastic achievement.
\newblock {\em The Quarterly journal of economics 114}, 2 (1999), 533--575.

\bibitem{aria2021comparison}
{\sc Aria, M., Cuccurullo, C., and Gnasso, A.}
\newblock A comparison among interpretative proposals for random forests.
\newblock {\em Machine Learning with Applications 6\/} (2021), 100094.

\bibitem{athey2019machine}
{\sc Athey, S., and Imbens, G.~W.}
\newblock Machine learning methods that economists should know about.
\newblock {\em Annual Review of Economics 11\/} (2019), 685--725.

\bibitem{brayne2017big}
{\sc Brayne, S.}
\newblock Big data surveillance: The case of policing.
\newblock {\em American sociological review 82}, 5 (2017), 977--1008.

\bibitem{card1993minimum}
{\sc Card, D., and Krueger, A.~B.}
\newblock Minimum wages and employment: A case study of the fast food industry in new jersey and pennsylvania, 1993.

\bibitem{district_attorney_cases}
{\sc {City and County of San Francisco}}.
\newblock District attorney cases prosecuted.
\newblock [Dataset], 2024.

\bibitem{comino2020silence}
{\sc Comino, S., Mastrobuoni, G., and Nicol{\`o}, A.}
\newblock Silence of the innocents: Undocumented immigrants’ underreporting of crime and their victimization.
\newblock {\em Journal of Policy Analysis and Management 39}, 4 (2020), 1214--1245.

\bibitem{police_department_2024}
{\sc {DataSF}}.
\newblock Police department incident reports: 2018 to present, 2024.
\newblock Online Database.

\bibitem{deschenes2017defensive}
{\sc Deschenes, O., Greenstone, M., and Shapiro, J.~S.}
\newblock Defensive investments and the demand for air quality: Evidence from the nox budget program.
\newblock {\em American Economic Review 107}, 10 (2017), 2958--2989.

\bibitem{efron2022computer}
{\sc Efron, B., and Hastie, T.}
\newblock Computer age statistical inference.

\bibitem{friedson2021shelter}
{\sc Friedson, A.~I., McNichols, D., Sabia, J.~J., and Dave, D.}
\newblock Shelter-in-place orders and public health: evidence from california during the covid-19 pandemic.
\newblock {\em Journal of Policy Analysis and Management 40}, 1 (2021), 258--283.

\bibitem{heckman1998matching}
{\sc Heckman, J.~J., Ichimura, H., and Todd, P.}
\newblock Matching as an econometric evaluation estimator.
\newblock {\em The review of economic studies 65}, 2 (1998), 261--294.

\bibitem{hernandez2019criminal}
{\sc Hern{\'a}ndez, W.}
\newblock Do criminal justice reforms reduce crime and perceived risk of crime? a quasi-experimental approach in peru.
\newblock {\em International review of law and economics 58\/} (2019), 89--100.

\bibitem{ioannidis2022end}
{\sc Ioannidis, J.~P.}
\newblock The end of the covid-19 pandemic.
\newblock {\em European journal of clinical investigation 52}, 6 (2022), e13782.

\bibitem{kahmann2022impact}
{\sc Kahmann, S., Hartman, E., Leap, J., and Brantingham, P.~J.}
\newblock Impact evaluation of the lapd community safety partnership.
\newblock {\em The Annals of Applied Statistics 16}, 2 (2022), 1215--1235.

\bibitem{lofstrom2020effect}
{\sc Lofstrom, M., Martin, B., and Raphael, S.}
\newblock Effect of sentencing reform on racial and ethnic disparities in involvement with the criminal justice system: The case of california's proposition 47.
\newblock {\em Criminology \& Public Policy 19}, 4 (2020), 1165--1207.

\bibitem{madansky1964instrumental}
{\sc Madansky, A.}
\newblock Instrumental variables in factor analysis.
\newblock {\em Psychometrika 29}, 2 (1964), 105--113.

\bibitem{maydeu2019instrumental}
{\sc Maydeu-Olivares, A., Shi, D., and Rosseel, Y.}
\newblock Instrumental variables two-stage least squares (2sls) vs. maximum likelihood structural equation modeling of causal effects in linear regression models.
\newblock {\em Structural Equation Modeling: A Multidisciplinary Journal 26}, 6 (2019), 876--892.

\bibitem{mchugh2013chi}
{\sc McHugh, M.~L.}
\newblock The chi-square test of independence.
\newblock {\em Biochemia medica 23}, 2 (2013), 143--149.

\bibitem{muhammad2019condemnation}
{\sc Muhammad, K.~G.}
\newblock {\em The condemnation of Blackness: Race, crime, and the making of modern urban America, with a new preface}.
\newblock Harvard University Press, 2019.

\bibitem{nordberg2002jails}
{\sc Nordberg, M.}
\newblock Jails not homes: Quality of life on the street of san francisco.
\newblock {\em Hastings Women's LJ 13\/} (2002), 261.

\bibitem{robinson2011evaluating}
{\sc Robinson, J.~P.}
\newblock Evaluating criteria for english learner reclassification: A causal-effects approach using a binding-score regression discontinuity design with instrumental variables.
\newblock {\em Educational Evaluation and Policy Analysis 33}, 3 (2011), 267--292.

\bibitem{wang2020deep}
{\sc Wang, S., Wang, Q., and Zhao, J.}
\newblock Deep neural networks for choice analysis: Extracting complete economic information for interpretation.
\newblock {\em Transportation Research Part C: Emerging Technologies 118\/} (2020), 102701.

\bibitem{zhou2021empirical}
{\sc Zhou, A., Koo, A., Kallus, N., Ropac, R., Peterson, R., Koppel, S., and Bergin, T.}
\newblock An empirical evaluation of the impact of new york's bail reform on crime using synthetic controls.
\newblock {\em arXiv preprint arXiv:2111.08664\/} (2021).

\end{thebibliography}

\clearpage
\section*{Tables}
\begin{table}[htbp]
  \centering
  \caption{Impact Of The Boudin Administration On SFDA Actions (Monthly, All Crimes)}
  \label{tab:regression_ca_ac}
  \begin{tabular}{lccccccc}
    \toprule
    & \multirow{1}{*}{(1)} & \multirow{1}{*}{(2)} & \multirow{1}{*}{(3)} & \multirow{1}{*}{(4)} \\
    \midrule
    \textit{Panel A. Log Monthly Total Prosecutions}  &   &   &  &  \\
    1. Boudin  & -0.1614*** & -0.3102*** & -0.1551*** & -0.3618***\\
     & (0.023) & (0.054) & (0.022) & (0.051) \\ \addlinespace

    \textit{Panel B. Log Monthly Total Convictions}  &   &   &  &  \\
    1. Boudin  & 0.0132 & -0.0301 & -0.2339*** & -0.2102*** \\
     & (0.095) & (0.083) & (0.048) & (0.061) \\ \addlinespace

    \textit{Panel C. Log Monthly Total Successful Diversions}  &   &   &  &  \\
    1. Boudin  & 0.817*** & 0.6909*** & 0.5784*** & 0.5845***\\
     & (0.086) & (0.105) & (0.058) & (0.076) \\ \addlinespace
     \textit{N}  & 120 & 120 & 108 & 108 \\
     Uses $Covid_1$ & X & & X & \\
     Uses $Covid_2$ & & X & & X \\
     With 2023 Excluded & & & X & X \\
    \bottomrule
  \end{tabular}

  \begin{flushleft}
    \textit{Notes:} (a) Standard Errors are heteroscedasticity robust using (White, 1980) covariance matrix; \\
    (b)*** $p<0.01$, ** $p<0.05$, * $p<0.1$
  \end{flushleft}
\end{table}
\begin{table}[htbp]
  \centering
  \caption{Impact Of The Boudin Administration On SFDA Actions (Monthly, Violent Crimes)}
  \label{tab:regression_ca_vc}
  \begin{tabular}{lccccccc}
    \toprule
    & \multirow{1}{*}{(1)} & \multirow{1}{*}{(2)} & \multirow{1}{*}{(3)} & \multirow{1}{*}{(4)} \\
    \midrule
    \textit{Panel A. Log Monthly Total Prosecutions}  &   &   &  &  \\
    1. Boudin  & -0.2352*** & -0.2947*** & -0.1551*** & -0.3618*** \\
     & (0.032) & (0.058) & (-0.022) & (-0.051) \\ \addlinespace

    \textit{Panel B. Log Monthly Total Convictions}  &   &   &  &  \\
    1. Boudin  & 0.0562 & 0.1511* & -0.1543*** & -0.0702 \\
     & (0.101) & (0.105) & (0.060) & (0.072) \\ \addlinespace

    \textit{Panel C. Log Monthly Total Successful Diversions}  &   &   &  &  \\
    1. Boudin  &  &  & 0.4931*** & 0.4725*** \\
     & & & (0.075) & (0.086) \\ \addlinespace
     \textit{N}  & 120 & 120 & 108 & 108 \\
     Uses $Covid_1$ & X & & X & \\
     Uses $Covid_2$ & & X & & X \\
     With 2023 Excluded & & & X & X \\
    \bottomrule
  \end{tabular}
  
  \begin{flushleft}
    \textit{Notes:} (a) Standard Errors are heteroscedasticity robust using (White, 1980) covariance matrix; \\
    (b)*** $p<0.01$, ** $p<0.05$, * $p<0.1$
  \end{flushleft}
\end{table}
\begin{table}[htbp]
  \centering
  \caption{Impact Of The Boudin Administration On SFDA Actions (Monthly, No Quality-of-Life Crimes)}
  \label{tab:regression_ca_noqol}
  \begin{tabular}{lccccccc}
    \toprule
    & \multirow{1}{*}{(1)} & \multirow{1}{*}{(2)} & \multirow{1}{*}{(3)} & \multirow{1}{*}{(4)} \\
    \midrule
    \textit{Panel A. Log Monthly Total Prosecutions}  &   &   &  &  \\
    1. Boudin  & -0.3625*** & -0.2545*** & -0.1551*** & -0.3618*** \\
     & (0.023) & (0.050) & (0.022) & (0.051) \\ \addlinespace

    \textit{Panel B. Log Monthly Total Convictions}  &   &   &  &  \\
    1. Boudin  & -0.3712*** & -0.0405 & -0.6101*** & -0.2138*** \\
     & (0.11) & (0.083) & (0.087) & (0.061) \\ \addlinespace

    \textit{Panel C. Log Monthly Total Successful Diversions}  &   &   &  &  \\
    1. Boudin  & 0.8674*** & 0.7844*** & 0.7118*** & 0.6868*** \\
     & (0.078) & (0.114) & (0.056) & (0.084) \\ \addlinespace
     \textit{N}  & 120 & 120 & 108 & 108 \\
     Uses $Covid_1$ & X & & X & \\
     Uses $Covid_2$ & & X & & X \\
     With 2023 Excluded & & & X & X \\
    \bottomrule
  \end{tabular}
  
  \begin{flushleft}
    \textit{Notes:} (a) Standard Errors are heteroscedasticity robust using (White, 1980) covariance matrix; \\
    (b)*** $p<0.01$, ** $p<0.05$, * $p<0.1$
  \end{flushleft}
\end{table}

\begin{table}[htbp]
  \centering
  \caption{Chi-Squared Test of Independence for Neighborhood Incident Report Concentrations}
  \label{tab:concentrations}
  \begin{tabular}{lcccc}
    \toprule
    & \multirow{1}{*}{$\chi^{2}$} & \multirow{1}{*}{$p$} & \multirow{1}{*}{$df$} & \multirow{1}{*}{Decision} \\
    \midrule
    1. All Crimes  & 0.014 & 1.0 & 40 & Fail to Reject Null\\

    2. Violent Crimes  & 0.017 & 1.0 & 40 & Fail to Reject Null \\
    \bottomrule
  \end{tabular}
\end{table}

\begin{table}[htbp]
  \centering
  \caption{Impact Of The Boudin Administration On SFPD Incident Reports (Monthly, Traditional 2SLS)}
  \label{tab:regression_incidents_std}
  \begin{tabular}{lccccccc}
    \toprule
    & \multirow{1}{*}{(1)} & \multirow{1}{*}{(2)} & \multirow{1}{*}{(3)} & \multirow{1}{*}{(4)} & \multirow{1}{*}{(5)} & \multirow{1}{*}{(6)} \\
    \midrule
    \textit{Panel A. First-Stage}  &   &   &  &  \\
    1. $\text{Boudin}_{t-1}$  & -0.1609*** & -0.3788*** & -0.1609*** & -0.3788*** & -0.1609*** & -0.3788*** \\
     & (0.025) & (0.070) & (0.025) & (0.070) & (0.0250) & (0.070)
 \\ \addlinespace

    \textit{Panel B. Second-Stage}  &   &   &  &  \\
    1. $\widehat{\text{Log Prosecutions}}_{t-1}$  & -0.3342*** & -0.0828 & -0.3018*** & -0.0795 & -0.1935*** & -0.01160 \\
     & (0.044) & (0.078) & (0.064) & (0.060) & (0.062) & (0.049) \\ \addlinespace
    \textit{N}  & 60 & 60 & 60 & 60 & 60 & 60 \\
     All Crimes & X & X & & & & \\
     Violent Crimes & & & X & X & & \\
     No QOL Crimes & & & & & X & X \\
     Uses $Covid_1$ & X & & X & & X \\
     Uses $Covid_2$ & & X & & X & & X \\
    \bottomrule
  \end{tabular}

  \begin{flushleft}
    \textit{Notes:} (a) Standard Errors are heteroscedasticity robust using (White, 1980) covariance matrix; \\
    (b) First-Stage equation calculated using Generalized-Least Squares regression; (c)*** $p<0.01$, ** $p<0.05$, * $p<0.1$
  \end{flushleft}
\end{table}
\begin{table}[htbp]
  \centering
  \caption{Impact Of The Boudin Administration On SFPD Incident Reports (Monthly, NN 2SLS)}
  \label{tab:regression_incidents_nn}
  \begin{tabular}{lccccccc}
    \toprule
    & \multirow{1}{*}{(1)} & \multirow{1}{*}{(2)} & \multirow{1}{*}{(3)} & \multirow{1}{*}{(4)} & \multirow{1}{*}{(5)} & \multirow{1}{*}{(6)} \\
    \midrule
    \textit{Panel A. Second-Stage}  &   &   &  &  \\
    1. $\widehat{\text{Log Prosecutions}}_{t-1}$  & -0.3337*** & -0.0595 & -0.3142*** & -0.0405 & -0.1837*** & -0.0003 \\
     & (0.043) & (0.074) & (0.065) & (0.058) & (0.060) & (0.045) \\ \addlinespace
     \textit{N}  & 60 & 60 & 60 & 60 & 60 & 60 \\
     All Crimes & X & X & & & & \\
     Violent Crimes & & & X & X & & \\
     No QOL Crimes & & & & & X & X \\
     Uses $Covid_1$ & X & & X & & X \\
     Uses $Covid_2$ & & X & & X & & X \\
    \bottomrule
  \end{tabular}

  \begin{flushleft}
    \textit{Notes:} (a) Second-Stage Standard Errors are heteroscedasticity robust using (White, 1980) covariance matrix; (b) First-Stage equation calculated using a Neural Network \textit{See Appendix For Neural Network Model Selection}; (c)*** $p<0.01$, ** $p<0.05$, * $p<0.1$
  \end{flushleft}
\end{table}
\begin{table}[htbp]
  \centering
  \caption{Impact Of The Boudin Administration On SFPD Incident Reports (Monthly, KNN 2SLS)}
  \label{tab:regression_incidents_knn}
  \begin{tabular}{lccccccc}
    \toprule
    & \multirow{1}{*}{(1)} & \multirow{1}{*}{(2)} & \multirow{1}{*}{(3)} & \multirow{1}{*}{(4)} & \multirow{1}{*}{(5)} & \multirow{1}{*}{(6)} \\
    \midrule
    \textit{Panel A. Second-Stage}  &   &   &  &  \\
    1. $\widehat{\text{Log Prosecutions}}_{t-1}$  & -0.3437*** & -0.0709 & -0.3163*** & -0.0792 & -0.2027*** & 0.0031 \\
     & (0.043) & (0.077) & (0.072) & (0.062) & (0.062) & (0.050) \\ \addlinespace
     \textit{N}  & 60 & 60 & 60 & 60 & 60 & 60 \\
     All Crimes & X & X & & & & \\
     Violent Crimes & & & X & X & & \\
     No QOL Crimes & & & & & X & X \\
     Uses $Covid_1$ & X & & X & & X \\
     Uses $Covid_2$ & & X & & X & & X \\
    \bottomrule
  \end{tabular}

  \begin{flushleft}
    \textit{Notes:} (a) Second-Stage Standard Errors are heteroscedasticity robust using (White, 1980) covariance matrix; (b) First-Stage equation calculated using a K-Nearest-Neighbors Algorithm \textit{See Appendix For Neural Network Model Selection}; (c)*** $p<0.01$, ** $p<0.05$, * $p<0.1$
  \end{flushleft}
\end{table}

\clearpage
\section*{Figures}
\begin{figure}[htbp]
  \centering
  \caption{Monthly Log Total Of SFDA Convictions}
  \begin{subfigure}[b]{0.48\textwidth}
    \includegraphics[width=\textwidth]{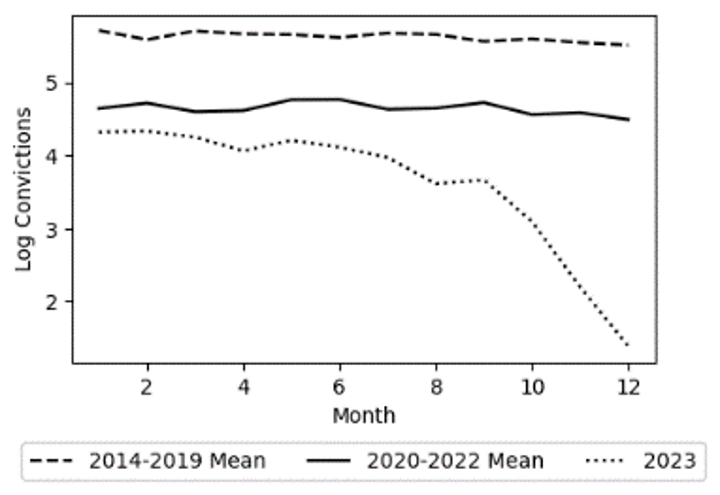}
    \caption*{Without Lags}
    \label{subfig:left_c}
  \end{subfigure}
  \hfill
  \begin{subfigure}[b]{0.48\textwidth}
    \includegraphics[width=\textwidth]{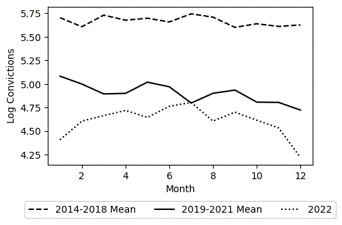}
    \caption*{With Lags \& 2023 Exclusive}
    \label{subfig:right_c}
  \end{subfigure}
  \label{fig:overall_c}
\end{figure}
\begin{figure}[htbp]
  \centering
  \caption{Monthly Log Total Of SFDA Successful Diversions}
  \begin{subfigure}[b]{0.48\textwidth}
    \includegraphics[width=\textwidth]{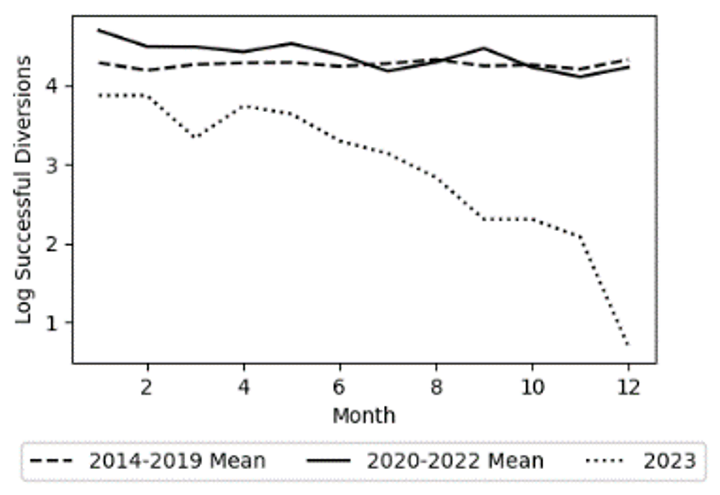}
    \caption*{Without Lags}
    \label{subfig:left_sd}
  \end{subfigure}
  \hfill
  \begin{subfigure}[b]{0.48\textwidth}
    \includegraphics[width=\textwidth]{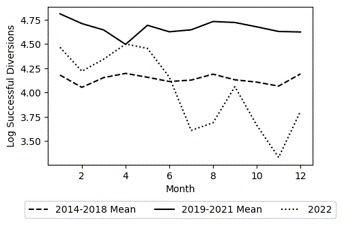}
    \caption*{With Lags \& 2023 Exclusive}
    \label{subfig:right_sd}
  \end{subfigure}
  \label{fig:overall_sd}
\end{figure}
\begin{figure}[htbp]
  \centering
  \caption{Monthly Log Total Of SFDA Prosecutions}
  \begin{subfigure}[b]{0.48\textwidth}
    \includegraphics[width=\textwidth]{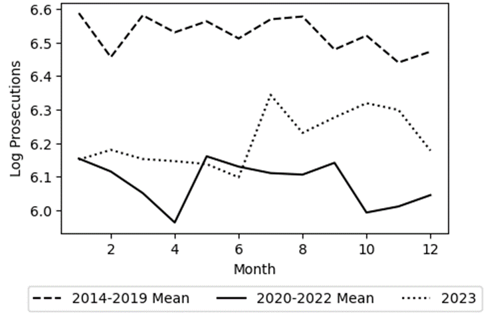}
    \caption*{Without Lags}
    \label{subfig:left_p}
  \end{subfigure}
  \hfill
  \begin{subfigure}[b]{0.48\textwidth}
    \includegraphics[width=\textwidth]{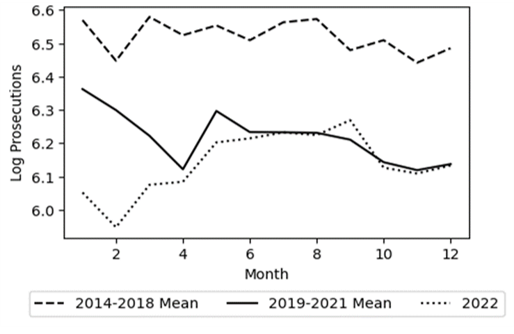}
    \caption*{With Lags \& 2023 Exclusive}
    \label{subfig:right_p}
  \end{subfigure}
  \label{fig:overall_p}
\end{figure}

\begin{figure}[htbp]
  \centering
  \caption{Yearly Incident Report Totals}
  \begin{subfigure}[b]{0.48\textwidth}
    \includegraphics[width=\textwidth]{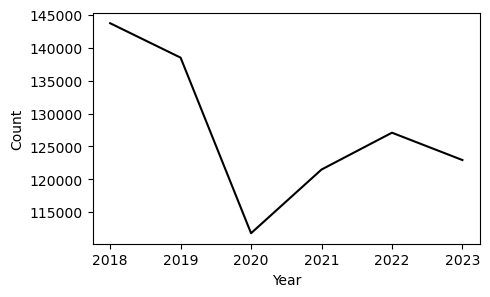}
    \caption*{All Crimes}
    \label{subfig:irt_ac}
  \end{subfigure}
  \hfill
  \begin{subfigure}[b]{0.48\textwidth}
    \includegraphics[width=\textwidth]{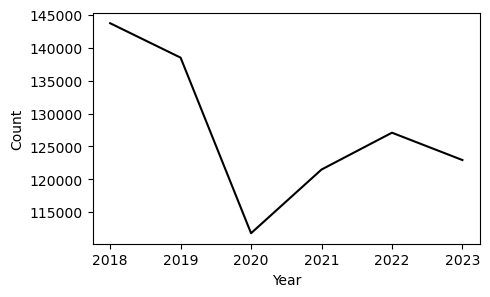}
    \caption*{Violent Crimes}
    \label{subfig:irt_vc}
  \end{subfigure}
  \label{fig:ir_tots}
\end{figure}
\begin{figure}
    \centering
    \caption{All Crimes Map (By Neighborhood)}
    \label{fig:grid_acm}
    
    \begin{subfigure}{0.42\textwidth}
        \includegraphics[width=\linewidth]{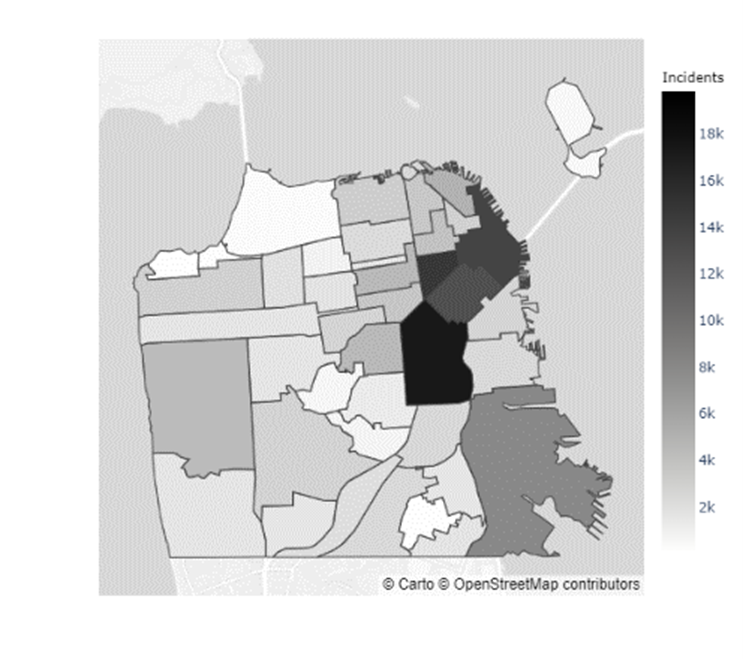}
        \caption*{2018}
        \label{fig:sub1_acm}
    \end{subfigure}
    \hfill
    \begin{subfigure}{0.42\textwidth}
        \includegraphics[width=\linewidth]{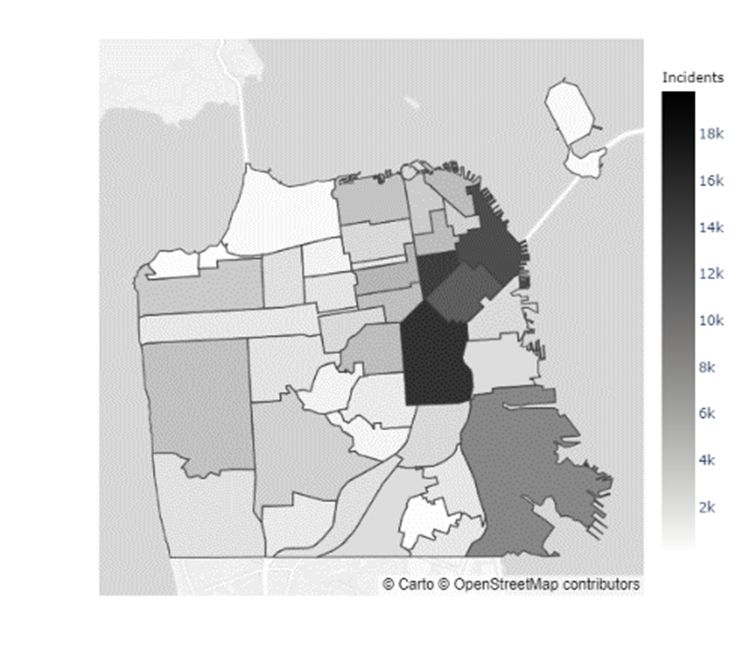}
        \caption*{2019}
        \label{fig:sub2_acm}
    \end{subfigure}

    \medskip

    \begin{subfigure}{0.42\textwidth}
        \includegraphics[width=\linewidth]{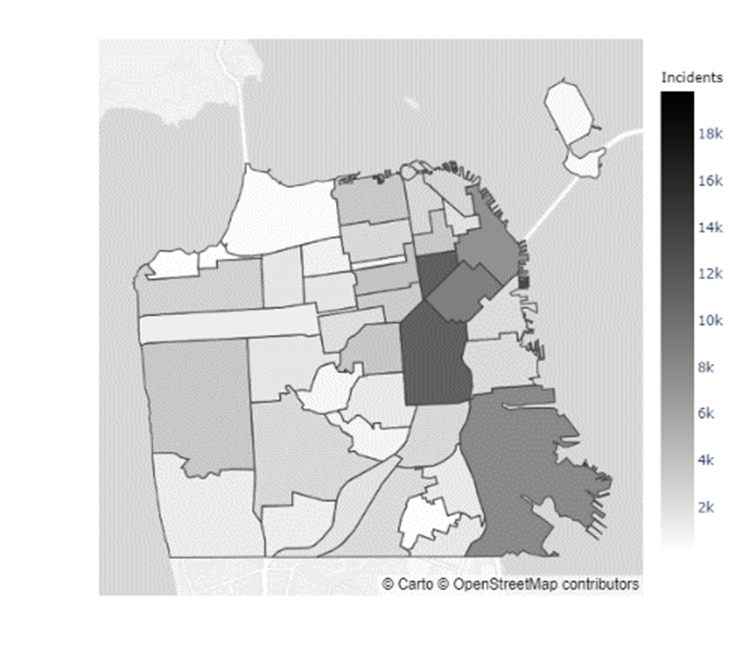}
        \caption*{2020}
        \label{fig:sub3_acm}
    \end{subfigure}
    \hfill
    \begin{subfigure}{0.42\textwidth}
        \includegraphics[width=\linewidth]{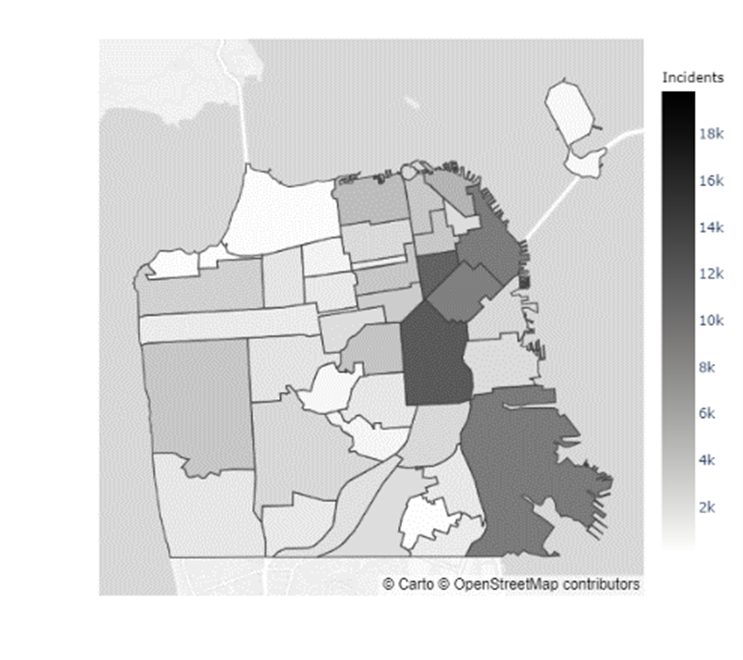}
        \caption*{2021}
        \label{fig:sub4_acm}
    \end{subfigure}

    \medskip

    \begin{subfigure}{0.42\textwidth}
        \includegraphics[width=\linewidth]{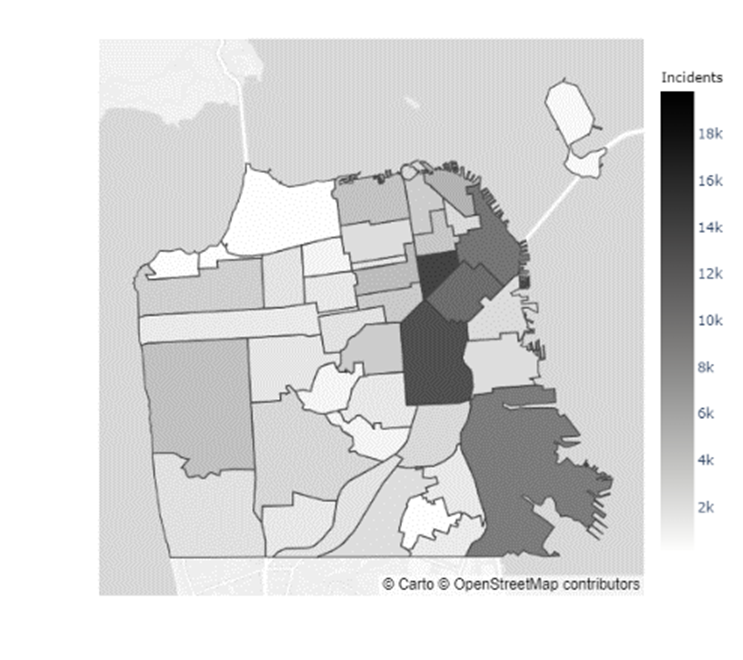}
        \caption*{2022}
        \label{fig:sub5_acm}
    \end{subfigure}
    \hfill
    \begin{subfigure}{0.42\textwidth}
        \includegraphics[width=\linewidth]{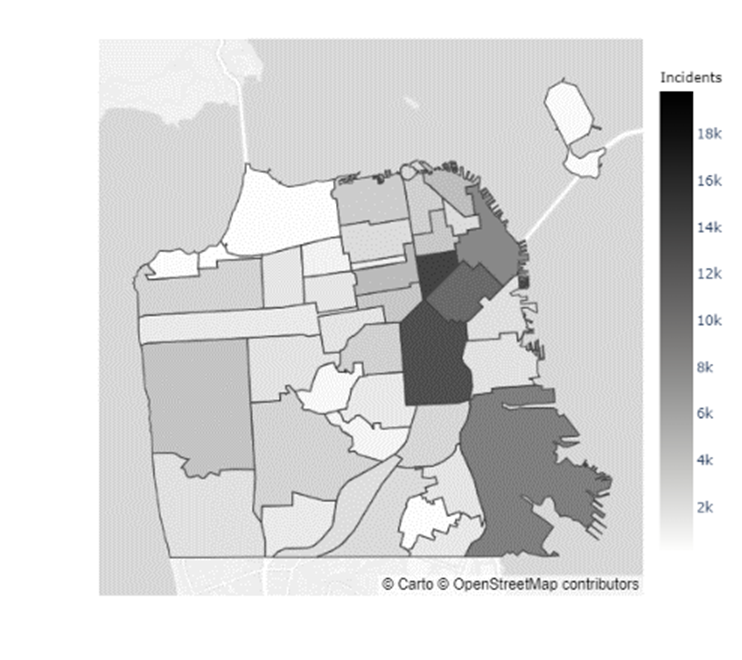}
        \caption*{2023}
        \label{fig:sub6_acm}
    \end{subfigure}
\end{figure}
\begin{figure}
    \centering
    \caption{Violent Crimes Map (By Neighborhood)}
    \label{fig:grid_vcm}
    
    \begin{subfigure}{0.42\textwidth}
        \includegraphics[width=\linewidth]{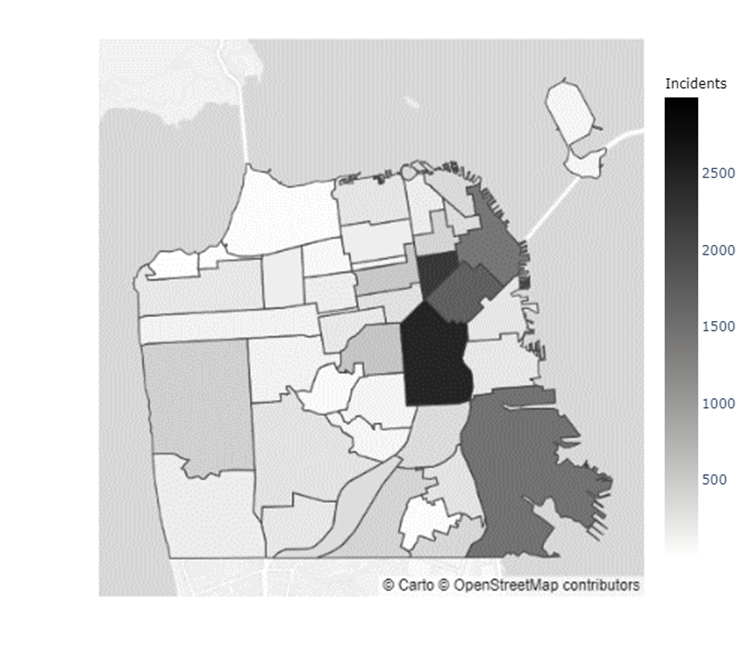}
        \caption*{2018}
        \label{fig:sub1_vcm}
    \end{subfigure}
    \hfill
    \begin{subfigure}{0.42\textwidth}
        \includegraphics[width=\linewidth]{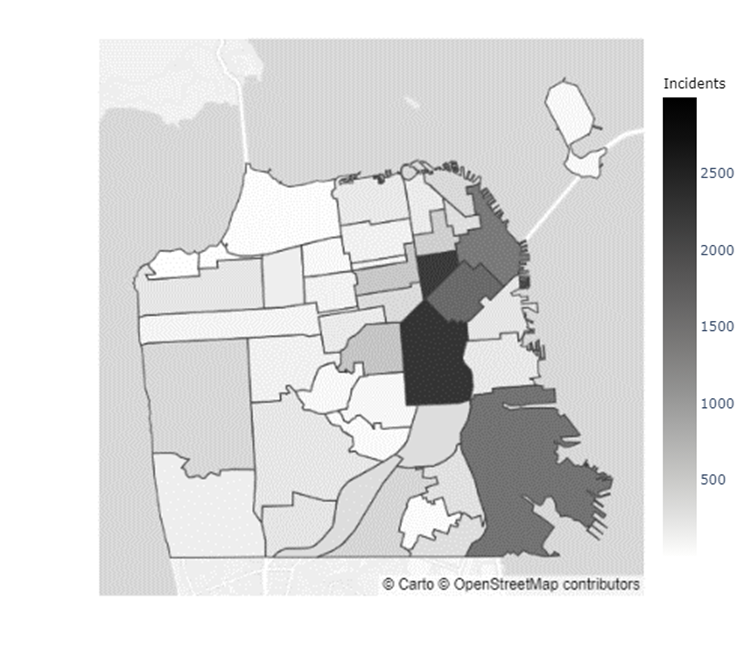}
        \caption*{2019}
        \label{fig:sub2_vcm}
    \end{subfigure}

    \medskip

    \begin{subfigure}{0.42\textwidth}
        \includegraphics[width=\linewidth]{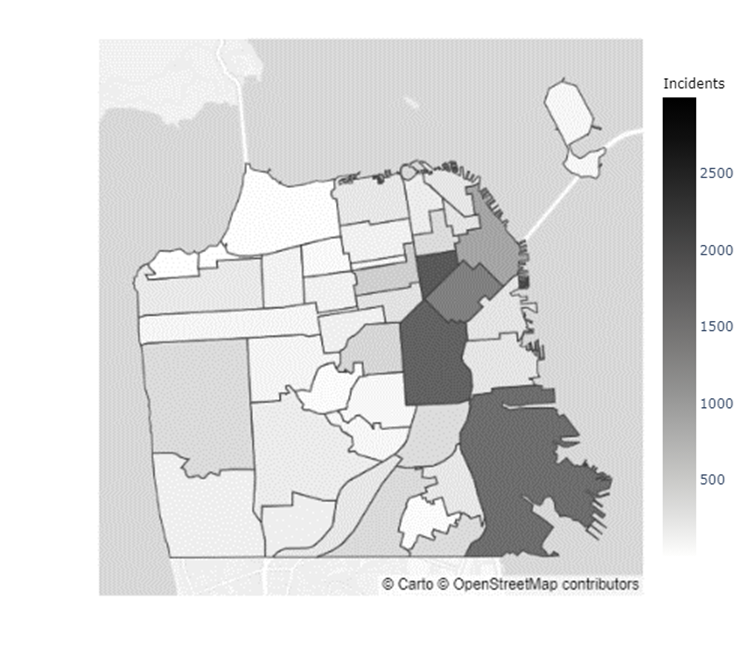}
        \caption*{2020}
        \label{fig:sub3_vcm}
    \end{subfigure}
    \hfill
    \begin{subfigure}{0.42\textwidth}
        \includegraphics[width=\linewidth]{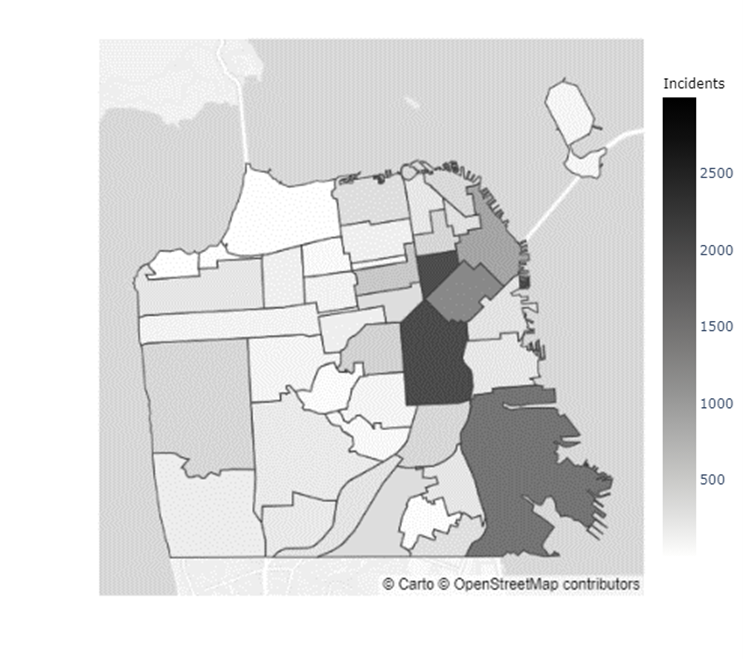}
        \caption*{2021}
        \label{fig:sub4_vcm}
    \end{subfigure}

    \medskip

    \begin{subfigure}{0.42\textwidth}
        \includegraphics[width=\linewidth]{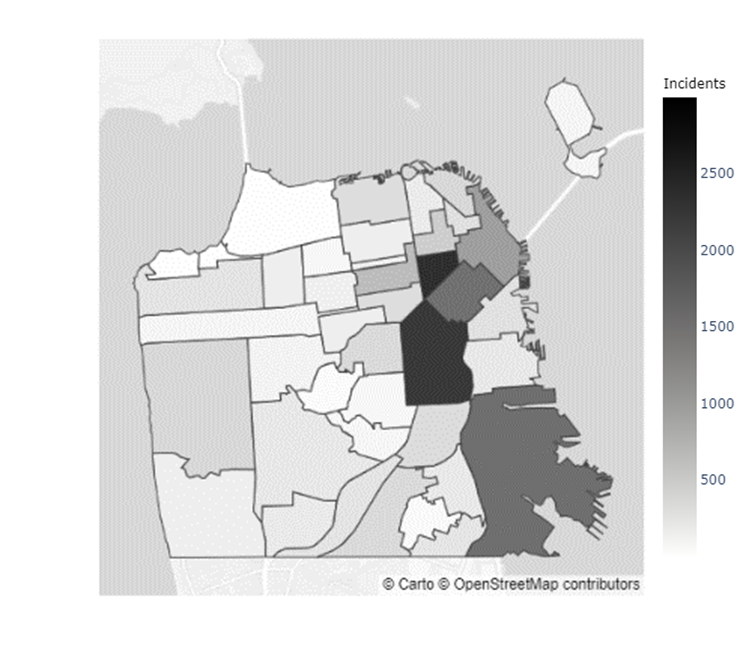}
        \caption*{2022}
        \label{fig:sub5_vcm}
    \end{subfigure}
    \hfill
    \begin{subfigure}{0.42\textwidth}
        \includegraphics[width=\linewidth]{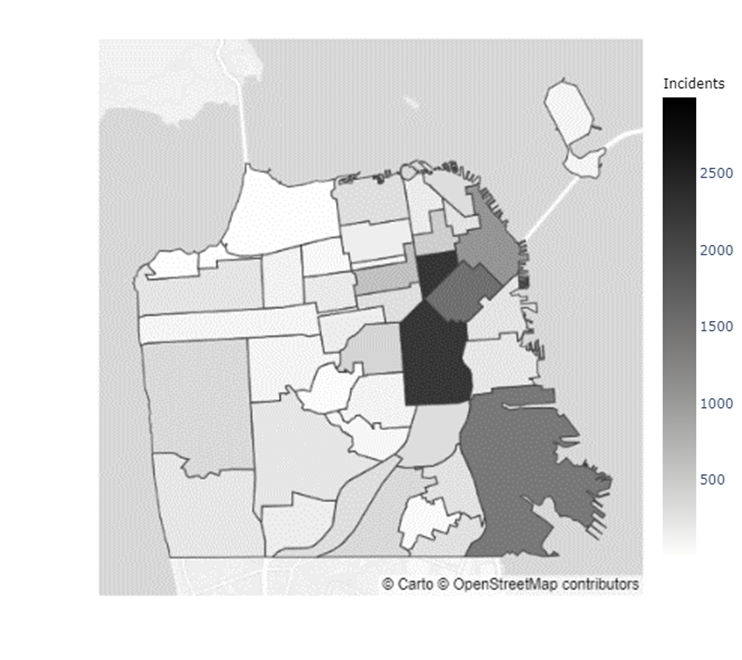}
        \caption*{2023}
        \label{fig:sub6_vcm}
    \end{subfigure}
\end{figure}

\clearpage
\appendix

\section{SFDA Case Resolution Summary Statistics}
\begin{table}[htbp]
  \centering
  \caption*{Table A1. All Crimes (Yearly)}
  \label{tab:cr_summary_stats}
  \resizebox{\textwidth}{!}{%
  \begin{tabular}{cccccccccc}
    \toprule
    \multirow{1}{*}{Arrest} & \multirow{1}{*}{} & \multirow{1}{*}{} & \multirow{1}{*}{Successful}  & \multirow{1}{*}{} & \multirow{1}{*}{Plead Guilty/} & \multirow{1}{*}{}  & \multirow{1}{*}{MATC} & \multirow{1}{*}{MATC} & \multirow{1}{*}{MATC} \\
    Year & Conviction & Dismissal & Diversion & Acquittal & Other & Total & (Overall) & (Conviction) & (Dismissal) \\
    \midrule
    2015 & 3460 & 934 & 526 & 36 & 508 & 5464 & 158 & 133 & 161 \\
    2016 & 3482 & 940 & 446 & 48 & 602 & 5518 & 173 & 157 & 175 \\
    2017 & 2858 & 1161 & 447 & 52 & 940 & 5458 & 199 & 181 & 202 \\
    2018 & 3327 & 1320 & 715 & 53 & 1054 & 6469 & 214 & 179 & 174 \\
    2019 & 3632 & 1246 & 1119 & 36 & 875 & 6908 & 219 & 194 & 219 \\
    2020 & 1976 & 879 & 1190 & 17 & 675 & 4737 & 323 & 285 & 301 \\
    2021 & 1258 & 703 & 1332 & 11 & 586 & 3890 & 421 & 310 & 283 \\
    2022 & 1397 & 690 & 1672 & 12 & 621 & 4392 & 403 & 327 & 259 \\
    \bottomrule
  \end{tabular}%
  }
  \begin{flushleft}
    \textit{Notes:} (a) MATC : Median Days from Arrest to Close
  \end{flushleft}
\end{table}

\begin{table}[htbp]
  \centering
  \caption*{Table A2. Violent Crimes (Yearly)}
  \label{tab:cr_summary_stats_vc}
  \resizebox{\textwidth}{!}{%
  \begin{tabular}{cccccccccc}
    \toprule
    \multirow{1}{*}{Arrest} & \multirow{1}{*}{} & \multirow{1}{*}{} & \multirow{1}{*}{Successful}  & \multirow{1}{*}{} & \multirow{1}{*}{Plead Guilty/} & \multirow{1}{*}{}  & \multirow{1}{*}{MATC} & \multirow{1}{*}{MATC} & \multirow{1}{*}{MATC} \\
    Year & Conviction & Dismissal & Diversion & Acquittal & Other & Total & (Overall) & (Conviction) & (Dismissal) \\
    \midrule
    2015 & 1039 & 283 & 104 & 23 & 86 & 1535 & 158 & 141 & 151 \\
    2016 & 972 & 284 & 98 & 26 & 102 & 1482 & 174 & 156 & 155 \\
    2017 & 860 & 339 & 131 & 35 & 166 & 1531 & 203 & 200 & 158 \\
    2018 & 995 & 411 & 214 & 29 & 132 & 1781 & 192 & 198 & 128 \\
    2019 & 1088 & 427 & 337 & 21 & 168 & 2041 & 266 & 239 & 190 \\
    2020 & 635 & 222 & 394 & 8 & 123 & 1382 & 409 & 406 & 258 \\
    2021 & 438 & 198 & 414 & 5 & 122 & 1177 & 519 & 478 & 370 \\
    2022 & 494 & 186 & 428 & 9 & 136 & 1253 & 428 & 394 & 365 \\
    \bottomrule
  \end{tabular}%
  }
  \begin{flushleft}
    \textit{Notes:} (a) MATC : Median Days from Arrest to Close
  \end{flushleft}
\end{table}

\clearpage
\section{Reduced Form Equation Model Performances}
\begin{table}[htbp]
  \centering
  \caption*{Table B: MSE's of Reduced Form Equations of Different 2SLS Models}
  \label{tab:first_stage_mses}
  \begin{tabular}{cccc}
    \toprule
    Type & Model & Specifics & MSE \\
    \midrule
    knn & ac\_c1 & [n\_neighbors: 12, weights: uniform] & 0.0032 \\
    nn & ac\_c1 & [hidden\_layer\_sizes: (16, 16, 8), activation: tanh, optimizer: sgd] & 0.0048 \\
    gls & ac\_c1 & - & 0.0053 \\
    gls & ac\_c2 & - & 0.0181 \\
    nn & ac\_c2 & [hidden\_layer\_sizes: (64, 32), activation: relu, optimizer: sgd] & 0.0185 \\
    knn & ac\_c2 & [n\_neighbors: 9, weights: uniform] & 0.0208 \\
    knn & noqol\_c1 & [n\_neighbors: 12, weights: uniform] & 0.0032 \\
    gls & noqol\_c1 & - & 0.0053 \\
    nn & noqol\_c1 & [hidden\_layer\_sizes: (16, 8), activation: tanh, optimizer: sgd] & 0.0053 \\
    nn & noqol\_c2 & [hidden\_layer\_sizes: (16, 8), activation: relu, optimizer: sgd] & 0.0145 \\
    gls & noqol\_c2 & - & 0.0181 \\
    knn & noqol\_c2 & [n\_neighbors: 9, weights: uniform] & 0.0208 \\
    knn & vc\_c1 & [n\_neighbors: 12, weights: uniform] & 0.0032 \\
    nn & vc\_c1 & [hidden\_layer\_sizes: (32, 32, 16), activation: tanh, optimizer: adam] & 0.0052 \\
    gls & vc\_c1 & - & 0.0053 \\
    nn & vc\_c2 & [hidden\_layer\_sizes: (32, 32, 16), activation: relu, optimizer: sgd] & 0.0172 \\
    gls & vc\_c2 & - & 0.0181 \\
    knn & vc\_c2 & [n\_neighbors: 9, weights: uniform] & 0.0208 \\
    \bottomrule
  \end{tabular}
  \begin{flushleft}
    \textit{Notes:} (a) ac = All Crimes, vc = Violent Crimes, noqol = No Quality-of-Life Crimes
    (b) c1 = $Covid_1$, c2 = $Covid_2$
  \end{flushleft}
\end{table}

\clearpage
\section{Map of San Francisco Neighborhoods}
\FloatBarrier
\begin{figure}[h]
    \centering
        \caption*{Figure C1.  Map of San Francisco Neighborhoods}
    \label{fig:sf_nhoods}
    \includegraphics[width=0.6\textwidth]{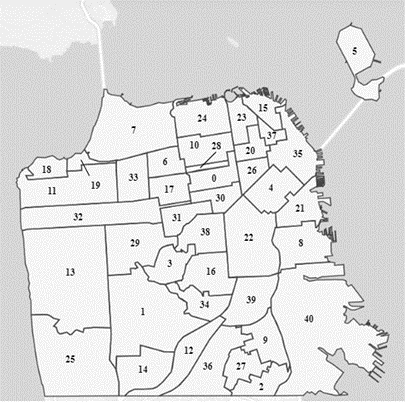}
\end{figure}
\begin{table}[h]
    \centering
    \caption*{}
    \label{tab:neighborhoods}
    \resizebox{\textwidth}{!}{%
    \begin{tabular}{cccccccc}
        \hline
        Index & Neighborhood & Index & Neighborhood & Index & Neighborhood & Index & Neighborhood \\
        \hline
        0 & Western Addition & 11 & Outer Richmond & 22 & Mission & 33 & Inner Richmond \\
        1 & West of Twin Peaks & 12 & Outer Mission & 23 & Russian Hill & 34 & Glen Park \\
        2 & Visitacion Valley & 13 & Sunset/Parkside & 24 & Marina & 35 & Financial District/South Beach \\
        3 & Twin Peaks & 14 & Oceanview/Merced/Ingleside & 25 & Lakeshore & 36 & Excelsior \\
        4 & South of Market & 15 & North Beach & 26 & Tenderloin & 37 & Chinatown \\
        5 & Treasure Island & 16 & Noe Valley & 27 & McLaren Park & 38 & Castro/Upper Market \\
        6 & Presidio Heights & 17 & Lone Mountain/USF & 28 & Japantown & 39 & Bernal Heights \\
        7 & Presidio & 18 & Lincoln Park & 29 & Inner Sunset & 40 & Bayview Hunters Point \\
        8 & Potrero Hill & 19 & Seacliff & 30 & Hayes Valley & & \\
        9 & Portola & 20 & Nob Hill & 31 & Haight Ashbury & & \\
        10 & Pacific Heights & 21 & Mission Bay & 32 & Golden Gate Park & & \\
        \hline
    \end{tabular}%
    }
\end{table}

\clearpage
\section{Reproducibility Documentation}

D1.1 \textit{Accessing the Repository}. In order to access the repository, clone it using the following code:

\begin{lstlisting}[language=bash, caption={}, label={lst:clonerepo}]
git clone https://github.com/jgte29/ucbhonorsthesis2023-2024.git
\end{lstlisting}

D1.2 \textit{Repository Structure}. See Next Page.

D2.1 \textit{Data Processing Pipeline -- Data Retrieval}. In my analysis, I depend on three distinct data sources: information on SFDA case actions, SFPD incident reports, and geospatial data for San Francisco neighborhoods.

D2.1.1 \textit{SFDA Case Actions}. The San Francisco case actions dataset is generated and overseen by the city's open data portal, DataSF. When the SFDA files charges against a defendant, relevant data is manually entered into the District Attorney Office's case management system. Approximately on a weekly basis, reports are extracted from this system, undergo cleaning and anonymization processes, and are subsequently integrated into the DataSF portal. Covering cases from January 2014 to the present, the dataset comprises information on over 100 thousand cases presented to the District Attorney's Office. It specifically includes cases where the office initiated prosecution by filing new criminal charges or submitting a motion to revoke probation or parole (MTR). Cases handled by the San Francisco Adult Probation Department or the state Division of Adult Parole Operations, which initiate a motion to revoke, are not included. In instances of new criminal charges, the most serious offense type is designated as the primary offense for prosecution by the SFDA. For MTR cases, the categorization is based on the initial prosecution, as filing an MTR indicates a pursuit of a new sanction within the sentence of a prior criminal conviction, rather than filing new charges for the latest offense. The filing date for MTRs is determined by the date when the SFDA filed the MTR for the new arrest.

D2.1.2 \textit{SFPD Incident Reports}. The dataset includes incident reports documented since January 1, 2018. These reports are recorded by law enforcement officers or through self-reporting by the public, utilizing the San Francisco Police Department's online reporting system. Reports are categorized based on the method of submission and the incident's nature. These categories encompass initial reports, signifying the first filing for a specific incident; coplogic reports, denoting online submissions by the public; and vehicle reports, covering incidents related to stolen and/or recovered vehicles. Subject to daily updates, the data is added to the open dataset only after undergoing review and approval by a supervising Sergeant or Lieutenant. Removal of incident reports from the dataset may occur due to compliance with court orders to seal records or for administrative reasons, such as ongoing internal affairs investigations or criminal inquiries.

D2.1.3 \textit{SF Neighborhoods}. Every entry in the San Francisco case actions dataset signifies a distinct action carried out by the district attorney’s office on a particular case involving a defendant. The temporal scope of my analysis is confined to crimes with arrest dates ranging from January 1, 2014, to December 31, 2023. Due to the dataset's detailed granularity, I aggregate the data on a monthly basis. The aggregations are conducted based on arrest dates rather than the dates of SFDA actions since the latter information is not available in the dataset.

D2.2.1 \textit{Data Processing Pipeline -- Exploratory Data Analysis: Temporal \& Spatial Aggregation}. \par

\noindent
\textit{SFDA Case Actions} Every entry in the San Francisco case actions dataset signifies a distinct action carried out by the district attorney’s office on a particular case involving a defendant. The temporal scope of my analysis is confined to crimes with arrest dates ranging from January 1, 2014, to December 31, 2023. Due to the dataset's detailed granularity, I aggregate the data on a monthly basis. The aggregations are conducted based on arrest dates rather than the dates of SFDA actions since the latter information is not available in the dataset.

\begin{figure}
\centering
\caption*{Table D1. The structure of the GitHub repository}
\framebox[\textwidth]{%
\begin{minipage}{0.9\textwidth}
  \dirtree{%
  .1 jgte29/ucbhonorsthesis2023-2024.
    .2 analysis.
    .3 ic\_data.
    .4 ica\_by\_month\_master.csv.
    .4 icv\_by\_month\_master.csv.
    .4 ic\_noqol\_by\_month\_master.csv.
    .4 incident\_counts\_a\_master.csv.
    .4 incident\_counts\_by\_crime\_master.csv.
    .4 incident\_counts\_v\_master.csv.
    .3 model\_performances.
    .4 first\_stage\_mses.csv.
    .4 ica\_c1\_knn\_model\_performance.csv.
    .4 ica\_c1\_nn\_model\_performance.csv.
    .4 ica\_c2\_knn\_model\_performance.csv.
    .4 ica\_c2\_nn\_model\_performance.csv.
    .4 icnoqol\_c1\_knn\_model\_performance.csv.
    .4 icnoqol\_c1\_nn\_model\_performance.csv.
    .4 icnoqol\_c2\_knn\_model\_performance.csv.
    .4 icnoqol\_c2\_nn\_model\_performance.csv.
    .4 icv\_c1\_knn\_model\_performance.csv.
    .4 icv\_c1\_nn\_model\_performance.csv.
    .4 icv\_c2\_knn\_model\_performance.csv.
    .4 icv\_c2\_nn\_model\_performance.csv.
    .3 regressions.
    .4 log\_conviction\_regressions.csv.
    .4 log\_incident\_tots\_regressions.csv.
    .4 log\_incident\_tots\_regressions\_kmeans.csv.
    .4 log\_incident\_tots\_regressions\_nn.csv.
    .4 log\_prosecutions\_lag1\_regressions.csv.
    .4 log\_prosecutions\_regressions.csv.
    .4 log\_successful\_diversion\_regressions.csv.
    .3 sf\_data.
    .4 Case\_Resoultions\_All\_Crimes.csv.
    .4 Case\_Resoultions\_Violent\_Crimes.csv.
    .4 DA\_Actions\_On\_Arrests\_All\_Crimes.csv.
    .4 DA\_Actions\_On\_Arrests\_Violent\_Crimes.csv.
    .4 District\_Attorney\_Cases\_Prosecuted.csv.
    .4 Outcomes\_of\_SFPD\_Incidents\_All\_Crimes.csv.
    .3 data\_analysis.ipynb.
    .2 Impact Analysis of the Chesa Boudin Administration.pdf.
    .2 README.md.
  }
\end{minipage}
}
\label{fig:package_github_struct}
\end{figure}

\noindent
\textit{SFPD Incident Reports} Similar to the detailed granularity of the SFDA case actions dataset, the SFPD incident reports dataset also necessitates monthly temporal aggregation. Moreover, for visualization purposes, I conduct yearly aggregations across various San Francisco neighborhoods. This approach allows me to depict the year-to-year trends in crime prevalence for each neighborhood.

D2.2.2 \textit{Data Processing Pipeline -- Exploratory Data Analysis: Feature Engineering}. \par

\noindent
\textit{SFDA Case Actions} In obtaining synthetic treatment and time-fixed effects for my analysis, I employed various methods. The development of the monthly time-fixed effect was straightforward, as it naturally emerged during the temporal aggregation process. However, acquiring the time-fixed effect associated with the COVID-19 pandemic and the indicator for Chesa Boudin being in office demanded a more nuanced approach. To account for the COVID-19 pandemic, I adopted two different strategies. The first involved a simple cutoff at March 2020, marking the commencement of Shelter-In-Place Orders in California. \protect\cite{friedson2021shelter} The second approach utilized a timeframe from March 2020 to the end of 2021, aligning with the literature suggesting the transition of COVID-19 to its endemic phase by the close of 2021. \protect\cite{ioannidis2022end} As SFDA action dates were unavailable in the case actions dataset, I implemented a lagging strategy to approximate when these actions occurred. Calculating the average Median Number of Days from Arrest to Conviction for the period from 2020 to 2022 revealed approximately 307 days (10 months) for all crimes and 426 days (14 months) for the subset of violent crimes. Consequently, for cases with arrest dates toward the end of Boudin's tenure, it was improbable that any conviction or successful diversion took place during his administration. Thus, the cutoffs for case actions under the Boudin administration are lagged by 10 and 14 months in the datasets for all crimes and violent crimes, respectively.

\noindent
\textit{SFPD Incident Reports} In incorporating month and COVID time-fixed effects, I apply a strategy parallel to the one used for the SFDA case actions dataset. However, for the synthetic treatment, I choose not to use lagging, adopting a similar approach to the one employed when analyzing prosecutions in the case actions dataset. 

\noindent
\textit{For more detailed information on the temporal aggregation and feature engineering for the datasets I use in my analysis, refer to the "Data" section of the paper.}

D3 \textit{Modeling}. I employ a regression discontinuity design to establish a causal connection between Chesa Boudin's tenure as SFDA and the occurrences of prosecutions, convictions, and successful diversions. Subsequently, I extend this quasi-experimental framework by leveraging Boudin's administration as an instrumental variable in exploring the effects of prosecutions on crime levels. \textit{For more detailed information on these models, refer to the "Methodology" section of the paper.}

D4 \textit{ML Hybrid Models}. In my study, I seek to harness the benefits of both deep learning algorithms and generalized least squares. By strategically incorporating a machine learning model, either k-nearest neighbors (knn) or neural networks, into the reduced form equation while maintaining the use of generalized least squares for the structural form, I aim to leverage the computational power of these algorithms to achieve reduced standard errors while preserving the interpretability inherent in traditional two-stage least squares (2SLS) methods. \textit{Refer to Appendix B of my paper for an evaluation of the machine learning algorithms' performance compared to GLS estimations in the reduced form equation.}

\end{document}